\PassOptionsToPackage{pdfa}{hyperref}
\documentclass[sigconf, nonacm]{acmart} 
\hypersetup{pdfapart=2,pdfaconformance=B}
\makeatletter
\def\hyxmp@xmptpg@schema{}
\makeatother
\usepackage[T1]{fontenc} 

\usepackage{pvldb}
\renewcommand\vldbdoi{10.14778/3849423.3849445}
\renewcommand\vldbpages{352 - 366}
\renewcommand\vldbissue{2}
\renewcommand\vldbyear{2026}
\renewcommand\vldbpagestyle{empty}
\renewcommand\vldbavailabilityurl{https://github.com/epfl-dias/proteus/tree/decom_op}

\usepackage{cleveref}
\usepackage{listings}
\usepackage{appendix}
\usepackage{longtable,booktabs,array}

\usepackage{tikz}
\usetikzlibrary{positioning,fit,arrows.meta,calc}

\usepackage{xcolor}
\usepackage{graphicx}
\usepackage{pifont}
\usepackage{xspace}
\usepackage{subcaption}

\newboolean{submission}
\setboolean{submission}{false}
\usepackage{multirow}
\usepackage{makecell}
\usepackage{tabularx}

\ifthenelse{\boolean{submission}}{
    \newcommand{\fixme}[1]{ }

}{
    \newcommand{\fixme}[1]{\textcolor{red}{FIXME: #1}}

}

\newcommand\ourOp{\texttt{CompressedBlockToTuples}\xspace}
\newcommand\ourOpAcro{\texttt{CB2T}\xspace}
\newcommand{\resolvemetadata}{\texttt{resolve\_metadata()}\xspace{}}
\newcommand{\decompresstile}{\texttt{decompress\_tile()}\xspace{}}
\newcommand\TtCB{\texttt{T2CB}\xspace}
\newcommand\tpb{\texttt{tpb}\xspace}
\newcommand\ipt{\texttt{ipt}\xspace}
\newcommand\BtT{\texttt{BlockToTuples}\xspace}
\newcommand\BtTAcro{\texttt{B2T}\xspace}
\newcommand\none{\textsc{Uncom}\xspace}
\newcommand\predecom{\textsc{Pre-decom}\xspace}
\newcommand\shared{\textsc{Shared}\xspace}
\newcommand\register{\textsc{Register}\xspace}
\newcommand\globalstage{\textsc{Global-stage}\xspace}

\definecolor{revcolor}{HTML}{D0021B}

\begin{document}

\immediate\pdfobj stream attr{/N 3} file{sRGB.icc}
\pdfcatalog{/OutputIntents [ <<
  /Type /OutputIntent
  /S /GTS_PDFA1
  /OutputConditionIdentifier (sRGB)
  /DestOutputProfile \the\pdflastobj\space 0 R
  /Info (sRGB)
>> ]}

\title{Tile-Based Decompression in Compiled GPU Engines}

\author{Antonio Boffa}
\email{antonio.boffa@epfl.ch}
\affiliation{%
  \institution{EPFL}
  \city{Lausanne}
  \country{Switzerland}
}

\author{Ananya Gupta}
\email{ananya.gupta@epfl.ch}
\affiliation{%
  \institution{EPFL}
  \city{Lausanne}
  \country{Switzerland}
}

\author{Hamish Nicholson}
\email{hamish.nicholson@epfl.ch}
\affiliation{%
  \institution{EPFL}
  \city{Lausanne}
  \country{Switzerland}
}

\author{Anastasia Ailamaki}
\email{anastasia.ailamaki@epfl.ch}
\affiliation{%
  \institution{EPFL}
  \city{Lausanne}
  \country{Switzerland}
}

\begin{abstract} 

Modern GPUs reach record analytical performance only when a query's working set fits within its GPU memory budget, and that budget is under growing pressure: GPUs are partitioned and shared, working sets are outgrowing high-bandwidth memory (HBM), and HBM is a fixed on-device resource whose per-dollar capacity has stagnated across generations. Lightweight compression can relieve this pressure by shrinking both footprint and bandwidth demand, but existing GPU compressed-execution techniques either bind decompression to a specific execution model or materialize decompressed columns in HBM before query execution.

We present a compression subsystem for compiled GPU query engines that treats decompression as a first-class physical operator. Compression-specific logic is isolated behind a narrow backend interface, so new GPU-oriented encodings can be added without rewriting downstream operators.

We implement the design in Proteus and study where decompressed values should be materialized in the GPU memory hierarchy. Our compiler-integrated shared-memory strategy fuses cooperative tile decompression with tuple-at-a-time query execution inside a single generated GPU kernel, avoiding decompressed-column materialization in HBM. Compared with HeavyDB and Sirius on GPU-resident data, fused on-chip decompression achieves competitive query times while reducing the resident memory footprint by up to 4$\times{}$ and memory-bandwidth consumption by up to 6$\times{}$.
\end{abstract}

\maketitle
\renewcommand\vldbdoi{10.14778/3849423.3849445}
\renewcommand\vldbpages{352 - 366}
\vldbtopmatter

\section{Introduction}\label{sec:intro}

Modern GPUs combine massive parallelism with multi-terabyte-per-second memory bandwidth, and recent GPU-native SQL engines show these hardware advantages can translate into record-setting end-to-end analytical performance~\cite{yu_2025, DBLP:conf/cidr/YogatamaYKSKCTP26}. This progress builds on more than a decade of GPU database research demonstrating that large speedups for core relational operators are achieved by carefully mapping them to GPU execution and memory behavior~\cite{Yuan:2013, Kaldewey:2012, 8731612, Paul:2021, ALCANTARA201239, 10.1145/3709689}. Achieving such performance, however, still depends critically on one condition: the working set must fit within the GPU memory budget.

\begin{figure}
\includegraphics[width=0.95\columnwidth]{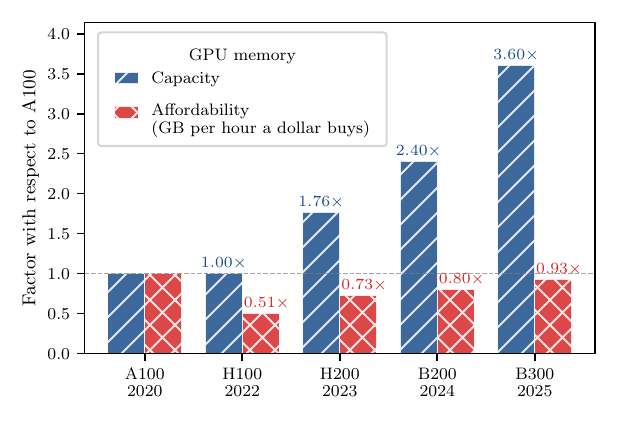}
\caption{Affordability gap between GPU memory capacity and the 
  memory each dollar buys. From A100 to B300, device memory grows by $3.6\times$, while the number of GB-hours of memory a dollar buys (averaged across four hyperscalers and one marketplace aggregator) sits below A100 levels. \\ Capacity scales; affordability does not.}\label{fig:capacity_vs_cost}
\end{figure}

\noindent{\bf Pressure on the GPU memory budget.}
This assumption is becoming increasingly fragile for several reasons. First, GPUs are increasingly shared across multiple workloads: mechanisms such as Multi-Instance GPU~\cite{nvidiamulti-instancegpu} and Multi-Process Service~\cite{multi-process} partition a single GPU into multiple isolated instances or processes, reducing the effective memory budget available to each job~\cite{DBLP:conf/cidr/JiangNSA26}. Second, the working sets themselves are growing faster than high-bandwidth memory (HBM) capacity: LLM serving is bottlenecked by KV-cache size and fragmentation~\cite{10.1145/3600006.3613165}, long-context training by activation storage~\cite{10.1145/3709703}, and analytical workloads usually operate over datasets that exceed device memory~\cite{10.14778/3749646.3749710}. 
Third, GPU memory is a fixed device resource: unlike CPU memory, which can often be expanded by adding DRAM, HBM capacity is determined by the GPU itself and cannot be increased without moving to a different device class or adding an entirely new GPU. 
Finally, memory pressure is also a cost-performance problem.

\noindent{\bf GPU memory price stagnation.}
Although recent hardware trends suggest that GPU memory is steadily increasing~\cite{DBLP:conf/cidr/YogatamaYKSKCTP26}, this view is incomplete once cost is considered. Based on our analysis across AWS~\cite{aws_calculator_2026}, Google Cloud~\cite{googlecalculator_2026}, Microsoft Azure~\cite{azure_calculator_2026}, Oracle Cloud~\cite{oracle_calculator_2022}, and marketplace aggregators~\cite{getdeploying_2026}, NVIDIA GPU device memory capacity has grown much faster than the per-dollar memory budget, opening a widening affordability gap. As shown in \Cref{fig:capacity_vs_cost}, from A100 to B300 device memory increases by $3.6\times$ (from $80$ to $288$~GB), while the number of GB-hours of memory a dollar buys is still only $0.93\times$ the A100 baseline (from $21$ to $19$~GB/h/\$), meaning a dollar today buys slightly less GPU memory than it did five years ago, despite over three times more memory shipping on each GPU. Larger HBM pools do not imply memory is cheap in practice; GPU memory remains a valuable and expensive resource, making memory-efficient execution critical for cost-performance. 

\noindent{\bf Data compression to the rescue.}
One approach to use as little GPU memory as possible while ensuring blazing-fast performance is lossless data compression, which reduces the volume of data that needs to be moved to and stored in precious HBM while preserving all information~\cite{Shanbhag:2022, 11271872, Huang:2025}. Further, as real-world datasets become increasingly redundant~\cite{DBLP:journals/csur/Navarro21a, DBLP:journals/pvldb/RenenHPVDNLSKK24, DBLP:journals/pvldb/ZengHSPMZ23, BOFFA2025112429}, compression becomes even more effective.

Compression, however, is not just a space occupancy optimization: once data is kept compressed in GPU memory, it fundamentally changes how query execution accesses and manipulates values. Compressed data is not directly consumable by traditional, compression-agnostic relational operators, and in practice, there is no single ``GPU compressed format'': different encodings offer different compression-ratio/decompression-speed trade-offs~\cite{Afroozeh:2024, Shanbhag:2022, 10.1145/3357384.3358067, Fang:2010, Li:2016, 10.1145/3736227.3736242}. 

\noindent{\bf Decompress-first vs execute-on-compressed data.}
On one hand, systems move compressed data into the GPU and then decompress it into a fixed-width uncompressed representation for downstream operators~\cite{Boeschen:2024, Nicholson:2025}. On the other hand, recent work targets executing operators directly on compressed representations, reducing both the HBM footprint and the HBM bandwidth spent reading/writing inflated values~\cite{Afroozeh:2024, 11271872, Huang:2025, 10.1145/3736227.3736242}.

The first line is largely driven by the host/storage-to-GPU transfer bottleneck, but GPU interconnect bandwidth is scaling rapidly (NVLink4 can reach 450 GB/s~\cite{nvidia_blackwell}). Also, materializing decompressed data into an uncompressed columnar form still incurs substantial intra-GPU overheads: it increases the HBM footprint and wastes HBM bandwidth by forcing inflated values to be written to and reread from global memory. We therefore focus on compressed execution to avoid materializing decoded columns in GPU memory, rather than only reducing the data transferred to the device.

\noindent{\bf GPU query execution models.}
 How compressed data can be consumed depends strongly on the underlying GPU execution model. At the coarse end, operator-at-a-time engines~\cite{10.1145/2882903.2882936, 10.1145/1620585.1620588, DBLP:conf/cidr/YogatamaYKSKCTP26, 10.14778/3551793.3551833} launch one kernel per operator over an entire column, materializing intermediate results, selection vectors, validity bitmaps, or mask tensors in HBM between operators. Refining this granularity, vectorized engines process columns in fixed-size contiguous batches called tiles~\cite{Shanbhag:2020, Shanbhag:2022}. GPU thread blocks cooperatively decode these tiles and retain cross-operator state on-chip rather than in HBM.
 
In contrast, compiled tuple-at-a-time engines generate pipelined GPU code in which each thread advances tuples through a fused operator sequence, aiming to minimize intermediate writes and keep values close to the consuming operators~\cite{Funke:2018, 10.14778/3275366.3284966, heavydb}. Proteus~\cite{DBLP:conf/cidr/RazaCSIAA20,Chrysogelos:2019,proteusweb,Nicholson:2025,DBLP:conf/cidr/NicholsonRCA23}, DogQC~\cite{Funke:2018, 10.14778/3380750.3380758}, Pyper~\cite{Paul:2020, Paul:2021}, HeavyDB~\cite{heavydb}, and Themis~\cite{10.14778/3705829.3705856} exemplify this direction while explicitly addressing SIMT challenges such as divergence and resource contention. With these optimizations, tuple-at-a-time execution can achieve up to $2.14\times$ shorter execution times than Crystal~\cite{Shanbhag:2020} (the fastest vectorized prototype engine~\cite{Cao:2023}).

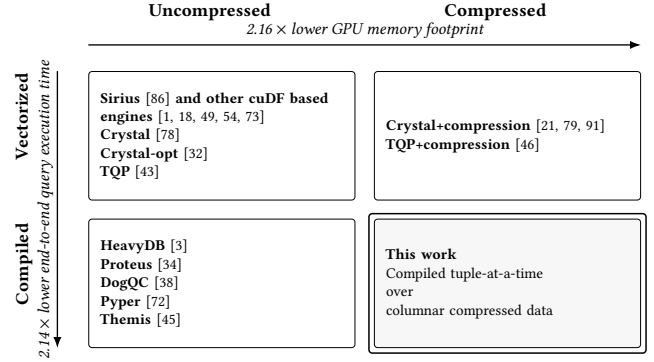
\begin{figure}
\centering
\resizebox{\columnwidth}{!}{%
  \begin{tikzpicture}[
  font=\normalsize,
  box/.style={
    draw,
    rounded corners=2pt,
    line width=0.5pt,
    inner sep=7pt,
    align=left,
    text width=5.1cm,   
    minimum height=2.7cm,
    anchor=north west
  },
  head/.style={font=\Large\bfseries},
  arrow/.style={-Latex, line width=0.8pt},
  note/.style={font=\normalsize\itshape, align=center},
]

\def\xL{0}
\def\xR{6.0}
\def\yTop{3.0}
\def\yBot{-0.1}

\node[head, anchor=south] at (\xL+2.55, \yTop+0.95) {Uncompressed};
\node[head, anchor=south] at (\xR+2.55, \yTop+0.95) {Compressed};

\node[box] (TL) at (\xL, \yTop) {%
\textbf{Sirius}~\cite{DBLP:conf/cidr/YogatamaYKSKCTP26} \textbf{and other cuDF based engines}~\cite{10.1145/3725324, kimball_natarajan_makkar_dice_haseeb, rapidsai_library_2026, 10.14778/3554821.3554829, blazingdb}\\
\textbf{Crystal}~\cite{Shanbhag:2020}\\
\textbf{Crystal-opt}~\cite{Cao:2023}\\
\textbf{TQP}~\cite{10.14778/3551793.3551833}\\
};

\node[box] (TR) at (\xR, \yTop) {%
\textbf{Crystal+compression}~\cite{Shanbhag:2022, 11271872, Afroozeh:2024}\\
\textbf{TQP+compression}~\cite{Huang:2025}\\
};

\node[box] (BL) at (\xL, \yBot) {%
\textbf{HeavyDB}~\cite{heavydb}\\
\textbf{Proteus}~\cite{Chrysogelos:2019}\\
\textbf{DogQC}~\cite{10.14778/3380750.3380758}\\
\textbf{Pyper}~\cite{Paul:2021}\\
\textbf{Themis}~\cite{10.14778/3705829.3705856}
};

\node[box,  fill=black!3] (BR) at (\xR, \yBot) {%
\textbf{This work}\\
Compiled tuple-at-a-time \\ over\\
columnar 
compressed data
};

\node[head, rotate=90, anchor=south] at (\xL-1.15, \yTop-0.95) 
{Vectorized};

\node[head, rotate=90, anchor=south] at (\xL-1.06, \yBot-0.95) 
{Compiled};

\draw[arrow] ($(TL.north west)+(0,0.55)$) -- ($(TR.north east)+(0,0.55)$);

\node[note] at ($(TL.north)!0.5!(TR.north)+(0,0.85)$) {\large{2.16$~\times$ lower GPU memory footprint}};

\draw[arrow] ($(TL.north west)+(-0.65,0)$) -- ($(BL.south west)+(-0.65,0)$);
\node[note, rotate=90] at ($(TL.west)!0.5!(BL.west)+(-0.9,0)$) {\large{2.14$~\times$  lower end-to-end query execution time}};

\node[draw, rounded corners=2pt, fit=(BR), line width=0.9pt] {};

\end{tikzpicture}%
}
\caption{Taxonomy of GPU query engines along two axes: execution model (vectorized vs.\ compiled) and data representation (uncompressed vs. compressed). The bottom-right quadrant highlights the gap we study: compiled tuple-at-a-time execution directly over lightweight compressed data. }
\label{fig:quadrant}
\end{figure}

\Cref{fig:quadrant} summarizes this landscape and motivates our focus on the unexplored quadrant: compiled GPU execution over compressed data, without eager intermediate-result materialization.

\noindent{\bf Decompression is tied to execution models.}
Although execute-on-compressed designs reduce HBM footprint, they typically make decompression reusable only within a specific execution model.
Crystal illustrates this trade-off clearly. Its tile-at-a-time programming model makes decompression reusable within a hand-written CUDA/C++ framework, but that reuse remains at the source-code level, with fixed execution structure and hard-coded parameters~\cite{Cao:2023}. It is not exposed as a compiler IR abstraction that can participate in whole-pipeline optimizations such as cross-operator inlining, explicit address-space placement, and specialization for registers versus shared memory~\cite{llvm-passes, nvvm}. As a result, the main benefits of compressed execution have so far been realized in specialized tile-oriented systems, not in compiled tuple-at-a-time engines.

\noindent{\bf Encoding-specific operators do not scale.} A second limitation is that recent execute-on-compressed designs embed format-specific decoding logic directly within relational operators (e.g., joins or aggregations) specialized for a particular encoding~\cite{Huang:2025, 11271872}. This creates an $N \times M$ problem: supporting $M$ encodings across $N$ operators requires $N \times M$ specialized implementations, reproducing the usual interoperability and maintenance bottleneck~\cite{10.14778/3749646.3749672}.
This tight coupling also creates an adoption problem~\cite{ 10.14778/3749646.3749672, 10.1145/3749163}. Although new compression techniques continue to improve compression ratio and decoding speed, they are difficult to deploy when support for each representation must be hard-wired into scans, operators, or storage readers. Existing formats, therefore, tend to ossify~\cite{kuiper_2025}: producers keep emitting older representations to remain compatible with existing engines, while newer compression techniques remain confined to prototypes or specialized systems.

\noindent{\bf A missing boundary in the compiled pipeline.}
The common issue behind these limitations is the absence of a stable boundary between compressed physical layouts and relational execution.
This is the same kind of modularity problem that has pushed data systems toward composable architectures, where monolithic DBMS stacks are decomposed into reusable layers with clean interfaces (for example, language frontends~\cite{ibis_2026}, optimizers~\cite{10.1145/3183713.3190662}, and execution engines~\cite{10.1145/3626246.3653368, 10.14778/3554821.3554829}) to better adapt to changing workloads, data formats, hardware, and deployment environments~\cite{10.14778/3603581.3603604, 10.14778/3749646.3749669, DBLP:conf/vldb/GavriilidisBPBQ22}. Following this trend, we argue that compressed GPU execution also needs a reusable, compiler-visible component boundary: new compression algorithms should require new metadata-resolution and tile-decompression logic, while the rest of the compiled relational pipeline remains unchanged.

\noindent{\bf Contributions:}
We design, implement, and evaluate a compression subsystem for a GPU-compiled query engine that operates directly on columnar data encoded with GPU-optimized lightweight compression schemes (GPU RFOR~\cite{Shanbhag:2022} and FastLanes BitPack~\cite{Afroozeh:2024}). Prior GPU compression work shows that tile decompression can be fused into hand-written query kernels. We lift this fusion into a tuple-at-a-time GPU query compiler by making decompression a first-class physical operator: compressed columns are decoded into the same tuple-binding interface used by uncompressed scans, allowing existing relational operators to consume them unchanged. This provides compiler-mediated fusion of GPU decompression and relational execution while preserving hardware-aware placement choices for decoded values. Our contributions are:

(i) We introduce \ourOp (\ourOpAcro), a physical data-flow decompression operator for compiled GPU query engines. It exposes columnar-oriented compressed columns as tuples, isolating compression-specific metadata resolution and tile loading behind a backend interface while preserving the relational operator interface (\Cref{sec:modular-decomp-op}).

(ii) \ourOpAcro enables compiler-mediated fusion of tile decompression with tuple-at-a-time GPU query pipelines. The generated code passes decoded values to downstream operators through global-memory, shared-memory, or register-backed bindings while preserving the standard tuple-at-a-time interface (\Cref{sec:compiler-scratchpad-decode}).

(iii) We integrate \ourOpAcro into Proteus and evaluate the runtime, memory traffic, and device-memory footprint trade-offs of uncompressed execution, eager decompression, staged global-memory, and fused on-chip decompression. Fusing decompression on-chip strictly dominates, delivering faster, smaller results on every query, while against uncompressed execution it roughly halves the resident footprint at a modest, query-dependent cost. 
Profiling further shows the best decode placement and launch geometry are query-dependent, and that peak occupancy does not always minimize runtime.
Across SSB~\cite{ONeil:2009}, TPC-H~\cite{tpch}, and Stack~\cite{10.1145/3448016.3452838} queries, we place these results in the broader GPU-SQL landscape, by comparing Proteus with \ourOpAcro against two state-of-the-art production GPU engines: HeavyDB~\cite{heavydb}, a compiled tuple-at-a-time engine, and Sirius~\cite{DBLP:conf/cidr/YogatamaYKSKCTP26}, a recent libcudf-based engine. Proteus with fused on-chip decompression achieves competitive query times while reducing the resident GPU memory footprint by up to 4$\times$ and GPU memory-bandwidth consumption by up to 6$\times$ (\Cref{sec:experiments}).

\begin{table*}[t]
\centering
\scriptsize
\setlength{\tabcolsep}{4.0pt}
\renewcommand{\arraystretch}{1.08}
\caption{Where compact or compressed representations enter GPU query execution. 
The distinction is whether compression is hidden behind a scan/decode boundary or exposed to downstream operator logic.}
\label{tab:compression-boundaries}
\begin{tabularx}{\textwidth}{@{}
p{0.23\textwidth}
p{0.09\textwidth}
p{0.195\textwidth}
p{0.285\textwidth}
p{0.13\textwidth}
@{}}
\toprule
\textbf{System / class} &
\textbf{Execution model} &
\textbf{Compression interface} &
\textbf{Operator contract} &
\textbf{Compression coupling} \\
\midrule

Sirius~\cite{DBLP:conf/cidr/YogatamaYKSKCTP26}; cuDF/Arrow engines &
Vector/op. &
Decode outside pipeline &
Operators consume decoded arrays &
Outside pipeline \\

Proteus~\cite{proteusweb,Chrysogelos:2019,DBLP:conf/cidr/RazaCSIAA20}; DogQC~\cite{10.14778/3380750.3380758}; Pyper~\cite{Paul:2021}; Themis~\cite{10.14778/3705829.3705856} &
Tuple/JIT &
Uncompressed scan/B2T &
Operators consume tuple bindings &
No compressed boundary \\

HeavyDB fixed-width storage~\cite{heavydb} &
Tuple/JIT &
Fixed-width column load &
Operators consume tuple bindings &
Type-narrowing load path \\

Crystal~\cite{Shanbhag:2020}; Crystal-opt~\cite{Cao:2023} &
Tile/AOT &
Uncompressed tile load &
Tile kernels consume staged values &
No compressed boundary \\

GPU-FOR/RFOR~\cite{Shanbhag:2022}; FastLanes-GPU~\cite{Afroozeh:2024} &
Crystal tile/vector &
Load/decode routine &
Tile kernels consume decoded values &
Codec-specific load path \\

GTileQuery~\cite{11271872} &
Crystal tile &
Direct compressed operators &
Operators inspect compressed metadata/values &
Operator--codec coupling \\

TQP~\cite{10.14778/3551793.3551833} &
Tensor/column &
Dense tensor representations &
Tensor operators consume dense tensors &
No compressed boundary \\

Compressed TQP~\cite{Huang:2025} &
Tensor/column &
Encoded tensor representations &
Operators rewritten over RLE/Index/Plain tensor layouts &
Encoding-aware operators \\

\textbf{CB2T in Proteus} &
\textbf{Tuple/JIT} &
\textbf{Compressed B2T physical operator} &
\textbf{Operators consume ordinary tuple bindings} &
\textbf{Backends only} \\

\bottomrule
\end{tabularx}
\label{tab:compression_boundaries}
\end{table*}

\looseness=-1
\section{Preliminaries and related works}\label{sec:related}

\subsection{GPU architecture essentials for database kernels}\label{ssec:gpu_essentials}

A GPU\footnote{We focus on NVIDIA GPUs. Other architectures, such as AMD GPUs, follow similar principles.} has many streaming multiprocessors (SMs). Threads are executed in groups of 32 called warps, and warps are scheduled on SMs. A thread block is a group of warps and is the basic unit of thread cooperation, synchronization, and shared-memory allocation.
Each thread has a private register file, which is the fastest but most limited storage level. 
Part of the on-chip memory in an SM can be configured as shared memory, explicitly managed by the programmer and visible to all threads in the same thread block. This allows frequently reused data to be staged on chip and reduces accesses to off-chip global memory. 
In CUDA, a kernel's execution configuration is fixed at launch time through \texttt{blockDim} (threads per block) and \texttt{gridDim} (blocks per grid). All thread blocks in the kernel share this configuration, and therefore all threads execute under the same launch parameters.
The GPU can place multiple thread blocks on the same SM, but only while respecting hardware limits. 
Increasing shared memory per block or registers per thread can therefore reduce the number of resident blocks and warps per SM, lowering occupancy and reducing the ability to hide latency.
Register pressure has two distinct effects. If the compiler cannot allocate all live values to registers, it emits spill stores and spill loads to thread-local memory. Although this memory is private to a thread, it is backed by the device memory hierarchy and is much slower than registers. 

\subsection{Compilation-based query execution and the GPU memory hierarchy}\label{ssec:rel:compilation}

Compilation-based engines~\cite{10.14778/2994509.2994516, heavydb} fuse relational pipelines by translating operator interactions into direct dataflow inside generated code, rather than materializing intermediates between operators. This idea follows the data-centric, produce/consume compilation model introduced by HyPer~\cite{Neumann:2011}, where values are kept in registers as long as possible. In this model, values loaded by a scan flow directly into predicates, hash computations, and aggregations within one fused code fragment. LLVM provides a single intermediate representation that can be lowered to different targets (e.g., x86 for CPUs and NVPTX for NVIDIA GPUs).

A scan operator can avoid exposing tuples through a generic runtime interface; instead, it can rely on input plug-ins~\cite{10.14778/2994509.2994516} that generate the access code needed for the specific storage format and place the required fields into virtual buffers. These buffers act as a tuple-level intermediate representation for the rest of the generated pipeline, and LLVM can often promote them to registers. 

\noindent{\bf From block-oriented storage to tuple: \BtT.}
We call \BtT (\BtTAcro) the component that bridges the uncompressed block-oriented storage layer and its tuple-oriented execution model: it iterates over the logical tuple positions in a block, exposing each as the tuple identifier and the requested attributes as ordinary tuple bindings.
For each position and field, it computes the element address with an LLVM \texttt{getelementptr} on the block base pointer and current index (issuing no memory access) and stores it in an \texttt{alloca} slot. LLVM's \texttt{mem2reg} pass then promotes these slots to SSA register values, leaving the per-element load emitted by the consuming operator as the only surviving memory access~\cite{llvm-passes}.
The identical IR is lowered to GPUs by the NVPTX backend without rewriting operator logic: the loop stride is set to the total thread count, so each thread processes a strided subset of positions, matching the SIMT model. For GPU targets, LLVM IR follows NVVM address-space conventions~\cite{nvvm}, where pointer types distinguish generic, global, shared, and local memory; these address spaces are preserved by NVPTX lowering and determine the corresponding PTX load/store instructions.

\looseness=-1
\subsection{GPU database execution models}\label{ssec:gpu_execution_models}

\Cref{tab:compression-boundaries} organizes the GPU query engines discussed in this section by their execution model and by where compression meets operator logic, and situates \ourOpAcro among them. We group these engines into three execution-model classes: tuple-at-a-time, vectorized, and tensor-at-a-time.

\noindent{\bf Tuple-at-a-time execution.}
A first line of work adopts a compil\-ation-based, tuple-at-a-time model, where multiple relational operators are fused into a small number of query-specific GPU kernels. 

Proteus is a GPU-accelerated columnar database that treats input formats through plug-ins and dynamically generates query-specific execution code, following a push-based, data-centric compilation model in which relational pipelines are fused into a single kernel (using LLVM~\cite{10.5555/977395.977673} and PTX~\cite{ptx}).
Funke et al.~\cite{Funke:2018} introduced fusion operators for GPUs, generating fused OpenCL/CUDA kernels that process tuples through data-parallel nested-loop pipelines. 
Subsequent work refined this model to match GPU execution constraints. 
He et al.~\cite{10.14778/3380750.3380758} proposed a JIT-compiled GPU query engine that mitigates warp divergence through lane refill and push-down operators, while Paul et al.~\cite{Paul:2021} introduced the Shuffle and Segment transforms to reduce divergence and resource contention in compiled GPU pipelines.
More recent work improves load balance by dynamically redistributing work across threads and warps~\cite{10.14778/3705829.3705856}. Optimizing warp utilization is orthogonal to our work; we integrate decompression into the same fused kernels via LLVM/NVVM.

HeavyDB~\cite{heavydb} (formerly OmniSci, formerly MapD), like Proteus, JIT-compiles fused query pipelines to GPU code through LLVM and PTX. HeavyDB
stores columns as contiguous arrays of fixed-width, byte-aligned values:
the address of row \(i\) is computed directly from \(i\) and the column's physical type width. This makes positional access simple and GPU-friendly, but
it also means that HeavyDB does not use variable-length lightweight encodings
such as RLE. The only data representation it allows is therefore
fixed-width integer narrowing, e.g., choosing an 8-, 16-, or 32-bit integer type for an entire column when the column range permits it.
  
We use HeavyDB in our experiments as a production-grade exemplar of the compiled tuple-at-a-time category, an independent cross-system baseline alongside Proteus; its restriction to inlined, fixed-width column loads isolates the rigidity that \ourOpAcro avoids by compressing at tile granularity.

\noindent{\bf Vectorized execution.}
A second line of work follows a vectorized processing model. On CPUs, this typically means that each operator processes a vector of columnar values at a time~\cite{boncz2005monetdb}. On GPUs, the same idea naturally extends to a tile-at-a-time model~\cite{Shanbhag:2020}, where a thread block cooperatively loads and processes a fixed-size tile in shared memory. 
Because a thread block loads cooperatively into shared memory and consumes the result within the same block, the tile-at-a-time model is where lightweight GPU compression has mostly been integrated, which we discuss in \Cref{ssec:lightweight_compression}. We omit Crystal-derived engines in our cross-system comparison: Crystal is not a SQL-driven system, since each SSB query is implemented as a hand-written CUDA kernel with hard-coded hash-table sizes, group counts, and output cardinalities, though it remains the reference for tile-at-a-time GPU execution. We discuss the benefit of \ourOpAcro with respect to Crystal-derived engines in \Cref{ssec:compressed-execution-baselines}.

Cao et al.~\cite{Cao:2023} analyze Crystal and other GPU DBMSs using roofline modeling and microarchitectural profiling, showing that kernel fusion, thread efficiency, and memory-system utilization remain central bottlenecks even in highly optimized GPU engines. 

Sirius~\cite{DBLP:conf/cidr/YogatamaYKSKCTP26} adopts a GPU-native, vectorized execution model built on top of libcudf. While it provides modular, drop-in acceleration via the Substrait plan format, its reliance on an uncompressed, Arrow-derived columnar format results in high HBM bandwidth consumption. 
Sirius treats decompression as an external pre-processing step and materializes uncompressed intermediate blocks between operator calls, missing the opportunity for the fused, compressed execution we propose. We test Sirius in our experiments as a state-of-the-art exemplification of the libcudf-based category.

\noindent{\bf Tensor-at-a-time execution.}
A third direction is to move GPU query execution into tensor runtimes. 
The Tensor Query Processor (TQP)~\cite{10.14778/3551793.3551833} compiles SQL into tensor programs executed by frameworks such as PyTorch, TVM, or ONNX Runtime, effectively treating relational pipelines as tensor computations scheduled by the ML stack. 
Conceptually, this is a batched execution model closer to tile-based processing than to compilation-based tuple-at-a-time engines.
 We do not benchmark TQP in our experiments: prior cross-system 
measurements~\cite{Cao:2023} report TQP at roughly $4\times$ HeavyDB's GPU-compute time on SSB, so the HeavyDB baseline already dominates TQP on the GPU-internal dimension our work targets.

\subsection{Lightweight GPU integer compression}\label{ssec:lightweight_compression}

Lightweight compression schemes, e.g, run-length, dictionary~\cite{Binnig:2009}, delta, bitmaps~\cite{Chambi:2016}, FSST~\cite{Boncz:2020, 10.1145/3719330.3721228}, XOR~\cite{Li:2005}, Null suppression~\cite{Roth:1993}, and learned schemes~\cite{Guerra:2025,Liu:2024, Boffa:2022} operate over blocks of column values~\cite{Abadi:2006}. They have been shown to be more effective than heavyweight compression schemes on GPUs~\cite{Nicholson:2025}. They are attractive for GPU analytics because they reduce data movement while keeping decoding simple enough to run close to query execution. GPU decompression is not only about choosing an encoding with a good compression ratio and/or decompression speed; the format must also expose sufficient parallelism for thousands of threads.

Early GPU compression schemes optimized individual lightweight primitives, such as null suppression and bit-packing~\cite{10.1145/3357384.3358067}. More recent designs move toward tile-based cascades, where multiple primitives are combined~\cite{Li:2016, Fang:2010}.  The primitives most relevant to this work are: bit-packing (BP), which stores each value using as few bits as possible, just enough to represent the column's value range; Frame-of-Reference (FOR), which subtracts a per-block minimum and bit-packs the residuals; and run-length encoding (RLE), which replaces a run of repeated values with a $(value, rl)$ pair. 
Most GPU systems that execute over lightweight encodings extend the vectorized, tile-at-a-time model of \Cref{ssec:gpu_execution_models}, and differ mainly in their encodings and in how tightly they couple decoding to operator logic. Shanbhag et al.~\cite{Shanbhag:2022} introduce GPU-RFOR, a compression-ratio-oriented scheme that cascades RLE, FOR, and BP: it encodes each tile as two streams (values and run lengths), each compressed with FOR. The cascaded design achieves higher compression ratios than single-layer encodings, but the multi-layer decode path is more compute-intensive. Its reference implementation is open source and exposes the decoder as GPU device code.

FastLanes-GPU~\cite{Afroozeh:2024} optimizes for decode throughput. Its FastLanes BP (FLBP) scheme applies FOR followed by BP in an interleaved layout that distributes values round-robin over SIMT lanes, so each thread in a warp decodes its share independently, without inter-lane transfers, yielding high bit-unpacking throughput. The trade-off is a lower compression ratio, since FLBP does not cascade RLE or other encodings (which have not been ported to CUDA).

Other tile-at-a-time designs push decoding past the codec and into operator logic itself. Zhang et al.~\cite{11271872} (GTileQuery) embed encoding-specific decoding directly within relational operators, so decoding runs as part of operator evaluation. Operating on the compressed representation directly avoids materializing decompressed columns in global memory and lowers memory traffic. The trade-off is generality: each operator is specialized to its encoding, coupling the storage format to the operator implementation.

Building on the tensor-at-a-time model (\cref{ssec:gpu_execution_models}), Huang et al.~\cite{Huang:2025} extend TQP with tensor-level primitives for filters, joins, and aggregations that operate directly on lightweight compressed formats. This keeps data compressed throughout execution but again intertwines physical encodings with runtime-specific operator implementations. Our goal is neither to design another GPU integer encoding nor another encoding-specialized relational operator.

\looseness=-1
\subsection{Challenges of compiled execution over compressed columns}
\label{sec:challenges}

A first challenge is a structural mismatch: lightweight encoding schemes align well with vectorized execution. A tile corresponds to a block of columnar values. Decompression can be implemented as a block-wide primitive applied to that tile. This explains why prior work integrates compression naturally into vectorized GPU engines~\cite{Shanbhag:2022}. Tuple-oriented, compiled execution, in contrast, operates on tuples via a fused pipeline~\cite{Neumann:2011}. 

This breaks the sequential, block-local access patterns that make column encodings efficient, increasing control divergence (variable-length codes, RLE runs, dictionary lookups), register pressure (multiple decode cursors/state), and memory traffic due to scattered reads across columns.

A second challenge follows from this. Since decoding one value at a time wastes the work shared across a tile, compressed columns are decoded cooperatively, a tile at a time. Their values must then be staged somewhere in the GPU memory hierarchy (registers, shared, or global memory) before the tuple-at-a-time pipeline reads them.
Unlike CPUs, where much of the cache hierarchy is transparent to the engine, GPU memory placement is explicit in the generated code: LLVM IR targeting NVPTX follows NVVM address-space conventions, where shared and global memory are distinguished through pointer types that lower to different PTX load/store instructions. The staging-location decision must therefore be exposed to address-space handling and register allocation, yet hidden from the operator interface. Otherwise, every operator would need to know how that column was materialized. 
\looseness=-1
\section{The modular decompression operator: a contract for composable compressed scans}
\label{sec:modular-decomp-op}

The design goal of \ourOp is to keep input data compressed in a columnar layout while feeding it to tuple-at-a-time relational operators within a compiled GPU query engine, without spilling decoded intermediates back to HBM. Doing so lowers the effective per-GB execution cost while keeping decompression composable. This section specifies the design (the contract \ourOpAcro commits to and why that boundary is the right one), and \Cref{sec:compiler-scratchpad-decode} realizes it as a JIT-compiled, fused GPU kernel.

\Cref{tab:compression-boundaries} summarizes how prior GPU query engines handle compressed data (detailed in \Cref{sec:related}): they are decompressed outside the pipeline into materialized intermediates, restricted to fixed-width type narrowing, or fused with execution but only by embedding compression-algorithm-specific logic inside relational or tile operators. \ourOpAcro is the only point in this space that stays tuple-at-a-time and JIT-compiled while confining all compression-algorithm knowledge to a reusable backend, so downstream operators consume ordinary tuple bindings regardless of the encoding, which is what we mean by a composable decompression operator.

Composability is well-understood for traditional relational operators and other engine components, but what composability means for decompression inside a compiled GPU query engine has not been characterized. We decompose composability for data compression into two concrete sub-properties:

\noindent(i) \textbf{Interface substitutability.} A compressed column presents the same logical attribute interface as an uncompressed column: relational operators consume values through the engine's standard scan/tuple interface rather than through compression-algorithm-specific APIs (\Cref{ssec:achieving-i}).

\noindent(ii) \textbf{Decompression encapsulation.} Compressed representation details, such as metadata layout, bit packing, run structure, and tile decoding, are confined to a reusable backend layer and do not appear in relational operator implementations. Therefore, the backend is responsible for resolving compressed-tile metadata and producing decoded tile values through the common interface; downstream operators remain oblivious to whether the values originated from raw memory, RFOR, bit-packing, run-length encoding, or another tile-decodable format (\Cref{ssec:achieving-ii}).

\subsection{Interface substitutability}
\label{ssec:operator-interface}
\label{ssec:achieving-i}

\begin{figure}
	\includegraphics[width=0.95\columnwidth]{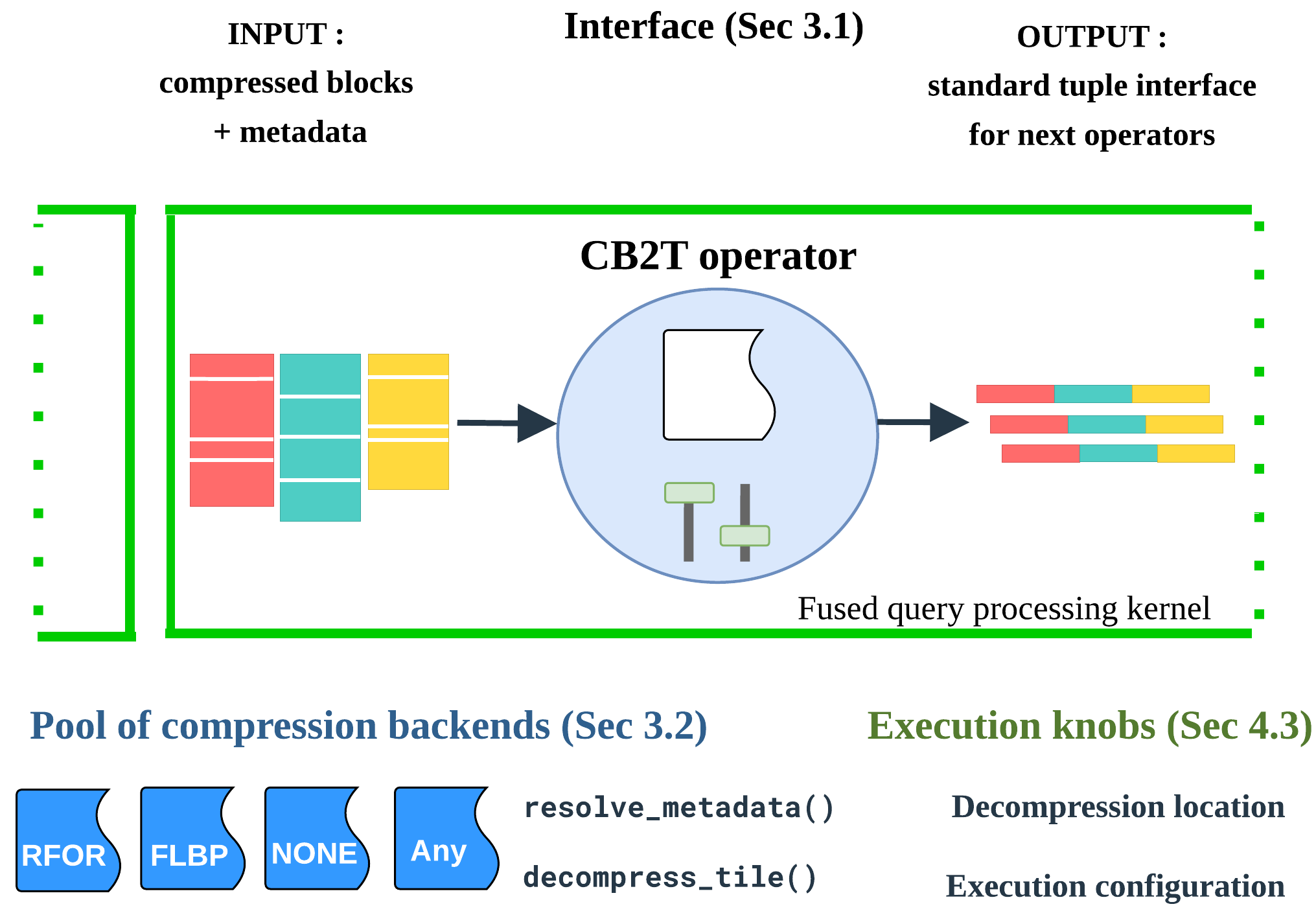}
	\caption{The modular decompression operator inside a fused query kernel: it consumes compressed tiles with per-tile metadata and emits a stable tuple stream to downstream operators (top), draws on interchangeable backends that implement only \resolvemetadata{} and \decompresstile{} (bottom-left), and is tuned by decode-placement and launch-configuration knobs (bottom-right).}\label{fig_logo}
\end{figure}

\Cref{fig_logo} summarizes the operator structure. From left to right, the operator receives compressed blocks that have already been moved to the GPU. A block is the unit of transfer and storage; internally it is partitioned into fixed-size tiles (the granularity at which the operator cooperatively decodes values and feeds them to the pipeline).
\ourOpAcro sits inside the fused query processing kernel and exposes two format-specific hooks, \resolvemetadata{} and \decompresstile{}, which are the subject of \Cref{ssec:achieving-ii}. In the figure's bottom-right corner, the orthogonal execution knobs control decode placement and launch configuration; these knobs are the levers exploited by \Cref{ssec:tile-size-knob}. Everything compression-algorithm-specific lives behind those two hooks; everything downstream of the operator boundary is fixed across compressed formats.

\ourOpAcro exposes a tuple-facing interface (the right-side arrow leaving the operator in \Cref{fig_logo}, labeled standard tuple interface for next operators) identical to the one a regular uncompressed scan would produce. Filter predicates, projections, hash-probes, and aggregations execute against ordinary tuple bindings and have no awareness of whether the values originated from a compressed or uncompressed source; their generated code is identical. 

\looseness=-1

\subsection{Decompression encapsulation}
\label{ssec:achieving-ii}

All compression-specific knowledge is confined to the backend layer shown in \Cref{fig_logo} as the Pool of compression backends, which contributes only two hooks: \resolvemetadata, which interprets the per-tile layout, and \decompresstile, which decodes one tile of values according to that layout. The two hooks follow the asymmetric cooperation model that is characteristic of lightweight GPU decompression: \resolvemetadata{} runs on a single block leader (one thread per block), which parses the per-tile header and prepares the fields, while \decompresstile{} executes cooperatively across all threads in the block, reading the metadata and emitting one decoded tile per column. 

This narrow backend contract also makes compressed format integration low-effort: adding a new one requires implementing only the two backend hooks for that format, how to interpret its metadata, and how to decompress one tile. This contract targets lightweight, tile-decodable column encodings, whose decode reduces to per-tile metadata resolution plus cooperative tile materialization~\cite{Shanbhag:2022, 11271872, Afroozeh:2024, xia_zhang_pan_liu_guan_zhang_du_2026}.
  
These hooks can be provided either as CUDA device functions or through the engine's LLVM IR generation interface, depending on whether the compressed format is implemented as hand-written GPU code or generated during query compilation. Once these two functions are registered, the new format automatically inherits the same logical column interface as existing encodings.
Compression algorithms are therefore free to evolve (new tile parameters, new header fields, even new variable-size sections) without touching any other component of the engine.

\noindent{\bf Self-describing blocks for engine-side isolation.}
Compression-related block metadata has to reach the decoder, but if it lived in tables maintained by the engine, it would force every component that touches a column to carry compression-aware state. \ourOpAcro eliminates this coupling by colocating all block metadata with the data it describes: each compressed block opens with a fixed header (format identifier, tile count, per-tile offsets), followed by the compressed payload (a sequence of tiles, each prefixed with its own metadata describing how to decode it). At scan time, the engine hands \ourOpAcro block pointers and nothing else. 
Nothing outside \ourOpAcro ever inspects compression-specific state: the rest of the engine treats compressed blocks as opaque byte ranges, indistinguishable from uncompressed ones, and routes them through the existing block-oriented path (allocation, transfer, eviction, mem-move) unchanged.

\noindent{\bf \ourOpAcro vs.\ data-access specialization.}
Input plug-ins~\cite{10.14778/2994509.2994516} specialize in how values are accessed from concrete dataset representations such as CSV, JSON, or relational binary layouts. They hide representation-specific details from operators by generating code that locates, parses, converts, or lazily reconstructs the values required by the query.

\ourOpAcro operates one layer later. It defines how that block is expanded into tuple bindings inside the fused pipeline. This is an execution problem rather than only an access problem: lightweight GPU decompression requires tile-level cooperation, metadata resolution, synchronization, and explicit placement of decoded values in shared memory or registers. In this sense, \ourOpAcro is a hardware-software co-designed component: it exposes the GPU structure needed for efficient decompression while keeping compression support composable.

\looseness=-1
\section{Realizing the contract: a JIT-compiled, fused GPU kernel}
\label{sec:compiler-scratchpad-decode}

\Cref{sec:modular-decomp-op} defines the operator interface and the format-specific backend contract that \ourOpAcro commits to at every site in the plan. This section shows how that contract is implemented inside a compiled GPU engine: each thread block first cooperatively decompresses a tile-sized row group of every compressed column into on-chip storage, then immediately invokes the next operator's tuple-at-a-time \texttt{consume()} code over the decoded tile, with both phases living in the same compiled kernel. Compilation is not part of the contract; it is the enabler that makes this implementation feasible. Because the operator boundary is represented inside the same LLVM IR module as the rest of the query, and lowered through the NVPTX backend following NVVM address-space conventions, the cooperative tile decode and the per-thread tuple pipeline can be fused into a single compiled kernel. The optimizer can then inline, propagate values, and eliminate redundant materialization across the decompression/operator boundary.

As \Cref{sec:exp-ssb-e2e} later shows, fused shared-memory decompression is the fastest compressed placement in every evaluated query--codec case. We therefore focus this section on the non-trivial system design choices required to realize the contract from Section~3 efficiently. \Cref{ssec:scratchpad-bridge} describes how we use shared-memory as a scratchpad, bridging cooperative decode and tuple-at-a-time execution; \Cref{ssec:tuple-binding-laundering} shows how the tuple binding handed to the next operator stays substitutable across the JIT boundary; \Cref{ssec:tile-size-knob} unpacks the execution knobs that govern decode placement and thread/tile configuration.

\subsection{Shared-memory as the bridge}
\label{ssec:scratchpad-bridge}

The mismatch between compressed input and tuple-at-a-time execution arises from mixing two access patterns: lightweight compression algorithms decode columnar values block-locally and cooperatively, with the entire thread block decoding as a single coordinated operation, exploiting the encoding's spatial locality and uniform structure. A tuple-at-a-time pipeline accesses values row-wise, in which each thread executes a fused sequence of operators.

Shared memory bridges the access pattern. Our operator emits a two-level loop: an outer loop assigns compressed tiles to thread blocks and invokes a cooperative decoder (encapsulated in the two hooks described in \Cref{ssec:achieving-ii}) that writes one tile per column into shared memory (using it as scratchpad); the inner tuple loop has each thread read its assigned values from the scratchpad, populate the tuple environment, and inline the \texttt{consume()} code of every downstream operator. Two \texttt{\_\_syncthreads()} barriers bracket the inner loop (one ensures the tile is fully decoded before any thread reads it, the other prevents overwrite while another thread is still consuming). Both phases live in the same compiled kernel, so decoded values flow from the decoder into the tuple loop through the shared scratchpad, without ever materializing in HBM.
\looseness=-1
\subsection{Preserving the tuple binding across the JIT boundary}
\label{ssec:tuple-binding-laundering}

As discussed in \Cref{ssec:rel:compilation}, \BtTAcro works because the data is already resident, uncompressed in global memory, and can be read directly. \ourOpAcro has no such option because it deals with compressed, variable-sized tiles and it decodes them in a cooperative, block-wide computation that must first reconstruct values and place them somewhere in the GPU memory hierarchy before the tuple-at-a-time pipeline can read them; registers, thread-block shared memory, and global memory each carry different costs in bandwidth, capacity, and occupancy, and the choice is observable in the IR because \Cref{ssec:rel:compilation} reminded us that the NVPTX backend treats pointers in different address spaces as distinct types. Interface substitutability (\Cref{ssec:achieving-i}) therefore demands that this placement decision terminate inside \ourOpAcro: the binding handed to the next operator must look identical to the one produced by \BtTAcro, regardless of where in the memory hierarchy the underlying value lives.               

\ourOpAcro meets this requirement with a two-sided interface around shared memory. Inside the operator, the cooperative decode routine is declared with shared-typed parameters and writes each column's decoded tile directly into shared memory. Externally, the same per-column scratchpad slot is exposed as a generic pointer, and that pointer is what enters the tuple binding consumed by the next operator. Downstream code issues a vanilla load on the generic pointer, unaware of where the value resides, and without a rewrite.

This encapsulation adds no indirection. Although the binding is exposed as a generic pointer, the backend can infer that it refers to the shared memory and emits a direct shared-memory access instead of a runtime-converted generic one, so downstream operators read each attribute directly from the on-chip tile. We confirmed this on the emitted PTX: shared memory accesses lower to \texttt{ld.shared}/\texttt{st.shared} with no address-space conversions, and the only remaining global loads are of the compressed input.

\subsection{Execution knobs: decode placement and launch configuration}
\label{ssec:tile-size-knob}

The shared-memory scratchpad of \Cref{ssec:scratchpad-bridge} is where \ourOpAcro places decoded values by default, but shared memory is not the only option. \ourOpAcro exposes two execution knobs that vary its physical realization without changing the external tuple binding:

\noindent{(i)} { \bf  Decode placement.}
We treat decode placement (where decoded values live between the cooperative decoder and the tuple-at-a-time loop) as an explicit design axis with three points: global memory, shared memory (\Cref{ssec:scratchpad-bridge}), and registers (\Cref{sec:exp-in-gpu-strategies}). The latter variant reduces shared-memory pressure by keeping only one decoded column tile in shared memory at a time and lifting each thread's assigned values into registers before reusing the scratchpad slot for the next column. These variants preserve the same external tuple binding; they differ only in where \ourOpAcro terminates the decode internally. For comparison, \Cref{sec:exp-in-gpu-strategies} also evaluates a global-memory materialization baseline, where decompression is performed by a separate kernel into an HBM buffer and the unchanged query pipeline later reads the decoded values.

\noindent{(ii)} { \bf Launch configuration.} The launch configuration determines how a fixed compressed tile is mapped to CUDA threads during decode and tuple consumption. The tile size, $T$, is the number of values handled per tile. This is a physical format parameter fixed when the dataset is compressed. At execution time, \ourOpAcro spreads the tile over a thread block: \texttt{threads\_per\_block} (\tpb) threads each decoding \texttt{items\_per\_thread} (\ipt) values, where $T = \tpb \times \ipt$.

\noindent{\bf  Occupancy as a resource tradeoff, not a bottleneck.}
Per-column shared tiles introduce a resource constraint absent from uncompressed scans. A fused compressed scan must reserve space for decoded values, metadata, and any compression-method-specific auxiliary state before the tuple-at-a-time loop consumes the tile. For $K$ columns and tile cardinality $T$, the decoded-value footprint alone scales with $K \times T$. This staging footprint can limit occupancy by reducing how many thread blocks fit concurrently on
an SM. \ourOpAcro therefore chooses occupancy as a tunable resource tradeoff rather than a fixed bottleneck. 

The first decision, the tile size $T$, is made when the data is compressed rather than at execution time. 
GPUs with larger shared-memory budgets can often afford larger tiles, while smaller shared-memory budgets or wider projections favor smaller tiles.

The second decision, made once $T$ is fixed by the compressed layout, is how to map the tile onto a thread block; \ourOpAcro exposes it through \tpb and \ipt. Larger \tpb (smaller \ipt) increases the number of resident warps per block and lowers per-thread register pressure, at the cost of more synchronization and setup overhead amortized over fewer values per thread; smaller \tpb (larger \ipt) does the reverse. Which end of this tradeoff is preferable is query- and codec-dependent and cannot be explained by occupancy alone, \Cref{sec:exp-ssb-e2e} quantifies it empirically.

\noindent{\bf Exposing the configuration matters because decode and query share one kernel.}
Because \ourOpAcro and the downstream relational operators execute inside the same GPU kernel, the decoder's cooperative shape is also the shape under which the fused query pipeline runs, while the query side competes for the remaining registers and cache. Joins and aggregations add probe and aggregation state on top of the shared memory already held for decoded tiles, so the configuration that maximizes decoder occupancy is rarely the one that minimizes end-to-end time. This is why \ourOpAcro exposes \tpb and \ipt as first-class knobs instead of hard-coding them: the same decoder can be retuned per query to balance its cooperative access pattern against the fused pipeline's footprint, without touching the decode routine or any relational operator. \Cref{sec:exp-ssb-e2e} quantifies this query-dependent sensitivity.
  
\looseness=-1
\section{SYSTEM INTEGRATION AND EVALUATION}
\label{sec:experiments}

We implement our \ourOpAcro and integrate efficient decompression into a composable, compilation-based, tuple-at-a-time GPU query engine.
We use Proteus as an example, which has the right infrastructure to study how decompression at a specific stage of query execution can affect query execution time~\cite{Nicholson:2025}.

Proteus explicitly abstracts the storage-to-pipeline transition through the \BtTAcro operator (\Cref{ssec:rel:compilation}). This allows us to implement \ourOpAcro (\Cref{sec:modular-decomp-op}) as a first-class, modular operator, capable of handling complex address-space transitions and collaborative decoding logic without modifying downstream relational operators. In this section, we answer three research questions:

\noindent{\bf RQ1} What is the end-to-end GPU memory-traffic cost of decompressing into a temporary GPU buffer prior to query execution, versus fusing it into the query pipeline? (\Cref{sec:exp:bw})
    
\noindent{\bf RQ2} Given this traffic analysis, what is the end-to-end query execution time of on-chip decompression, and is it always beneficial to fuse decompression and query execution given its shared-memory and register pressure? (\Cref{sec:exp-ssb-e2e})

\noindent{\bf RQ3}  
How does fused on-chip decompression in Proteus compare in end-to-end runtime and GPU-memory footprint against state-of-the-art GPU engines: HeavyDB and Sirius? (\Cref{ssec:competitors})

\noindent{\bf Hardware.} 
We ran the complete evaluation on both a local NVIDIA A40 GPU with 48~GB of GDDR6 memory and a cloud NVIDIA A100-SXM4 GPU with 40~GB of HBM2 memory, obtaining consistent qualitative conclusions and similar relative performance trends. We report the A100 results because its HBM memory system more closely matches the hardware setting motivating this work (\Cref{sec:intro}). The A100 instance was provided by lambda.ai~\cite{lambda, DBLP:conf/cidr/YogatamaYKSKCTP26} and runs Ubuntu~22.04.5 with Linux kernel~6.8.0. The instance exposes 30 CPU cores of an AMD EPYC~7J13 processor and 216~GiB of DRAM. 
Each system is built and run with its system-specific toolchain. Proteus is built with CUDA~11.3 and LLVM/Clang~14.0. HeavyDB was built with CUDA~11.8, GCC~12.4, and LLVM/Clang~14.0. Sirius uses DuckDB~v1.5.2, CUDA~12.9, and GCC~14.3.
  
\noindent{\bf Dataset.}
We use the SSB dataset SF 100, selected TPC-H queries SF 100, and real-world Stack queries over the fixed StackExchange dump.
Following standard practice in GPU databases~\cite{Shanbhag:2020, Shanbhag:2022}, for string columns, regardless of the compression scheme used, we apply dictionary encoding during data preparation, uniformly across all tested systems. Each unique string value is mapped to a 4-byte integer ID using a static ordered dictionary constructed from the full column. 
Each column is partitioned into fixed-size 2MiB blocks: the granularity Proteus uses to overlap host-to-device transfer with processing; our in-GPU-resident experiments preload all blocks and exclude transfer, but keep this block size for fidelity with the general data path~\cite{Boeschen:2024, Nicholson:2025} (see \Cref{sec:in-gpu-methodology}). 
In the compressed dataset, each block is further divided into 1024-value tiles. In \Cref{sec:exp-ssb-e2e}, we decrease RFOR tile size to $T{=}512$ since it is more tile-size–sensitive than FLBP because each tile stages more shared memory (run-lengths and cooperative scan buffers). 

\noindent{\bf Workload.}
We run the SSB queries Q1.1/2.1/3.1 that scan 9.6 GB and Q4.1 that scans 14.4 GB per query. We additionally report TPC-H Q5 and Q12, whose uncompressed referenced-column footprints are 11.5 and 13.2~GB, respectively. Both join \texttt{lineitem} and \texttt{orders}, the two largest TPC-H relations. Finally, among the Stack workload, we select Q11 as a representative multiway join and Q14 as the deepest join plan, spanning four to seven relations and three to six hash joins over real-world data. Cross-system results on these workloads should be interpreted as production-baseline comparisons: the systems differ in memory-pool policy, type lowering, SQL rewriting, and join planning. Building on studies that characterize GPU query engines across systems and hardware~\cite{Cao:2023, Kabic:2025} and on single-system comparisons of competing execution designs~\cite{10.14778/3275366.3284966}, future work will compare across more workloads, execution paradigms, hardware, and data representations.

\subsection{\bf In-GPU evaluation methodology}
\label{sec:in-gpu-methodology}

Proteus was designed for heterogeneous CPU-GPU execution. It moves data as per-column 2MiB blocks; once a row group's blocks are all resident, it launches the compiled pipeline over the row group, preserves state across launches, and continues to the next.

Here we switch to a single-kernel-per-pipeline execution mode: each compiled pipeline runs as a single kernel launch over the entire input (as in HeavyDB). All input data are already resident in GPU global memory, and this mode lets us isolate the impact of decompression location within the GPU memory hierarchy. 
On the uncompressed baseline, single-kernel-per-pipeline reduces query time by 19\% on average, demonstrating that for in-GPU execution, the per-launch CPU-GPU control-plane cost is itself a first-order term once data movement is removed. 

\subsection{Compression algorithms}

In all experiments, the same compression algorithm is applied to every column and block. Strategies that select different compression algorithms per column or per block (potentially improving compression ratios by matching the encoding to each column's value distribution~\cite{10.14778/3503585.3503588}) are left for future work, along with an analysis of the additional query compilation time~\cite{DBLP:conf/sigmod/LangMFB0K16}. 

Inside the \ourOpAcro operator, we evaluate two open-source, lightweight integer compression schemes: RFOR and FLBP (introduced in \Cref{ssec:lightweight_compression}), which represent different points in the design space (as shown in \Cref{tab:decomp_per_query}): one optimized for compression ratio via cascaded encoding layers, and one optimized for raw decompression speed via a simpler, fully data-parallel layout.

We define the compression ratio as the ratio of the uncompressed representation to the compressed representation. We report compression ratios and decompression speed for NVIDIA nvCOMP Cascaded~\cite{nvcomp} as a reference point, but do not integrate it as a query operator: nvCOMP is a closed-source library whose decompression routines are not available as GPU device code, making it impossible to inline into the LLVM IR compilation stack and fuse with downstream operators within the same kernel; furthermore, it is outperformed by schemes with fused, single-kernel decoders~\cite{Shanbhag:2022,Afroozeh:2024,10.1145/3736227.3736242}. 

\begin{table}[t]
\centering
\caption{%
  Compression ratio and aggregate GPU decompression throughput
  (GB/s) over each query's columns.
}
\label{tab:decomp_per_query}
\begingroup
\scriptsize
\renewcommand{\arraystretch}{0.95}
\begin{tabular*}{\columnwidth}{@{\extracolsep{\fill}}l rr rr rr@{}}
\toprule
& \multicolumn{2}{c}{\textbf{FLBP}}
& \multicolumn{2}{c}{\textbf{RFOR}}
& \multicolumn{2}{c}{\textbf{nvCOMP Casc.}} \\
\cmidrule(lr){2-3}\cmidrule(lr){4-5}\cmidrule(lr){6-7}
\textbf{Query}
  & Ratio & GB/s
  & Ratio & GB/s
  & Ratio & GB/s \\
\midrule
SSB Q1.1   & 2.45 & \textbf{713} & \textbf{2.96} & 301 & 2.64 & 135 \\
SSB Q2.1   & 1.59 & \textbf{576} & \textbf{1.82} & 282 & 1.80 & 138 \\
SSB Q3.1   & 1.52 & \textbf{559} & \textbf{2.30} & 296 & 2.26 & 160 \\
SSB Q4.1   & 1.54 & \textbf{566} & \textbf{2.02} & 286 & 2.01 & 145 \\
TPC-H Q5   & 2.08 & \textbf{653} & 2.19 & 284 & \textbf{2.25} & 136 \\
TPC-H Q12  & 2.57 & \textbf{729} & \textbf{2.74} & 271 & 2.58 & 131 \\
Stack Q11  & 3.54 & \textbf{625} & \textbf{4.18} & 313 & 4.12 & 125 \\
Stack Q14  & 3.19 & \textbf{646} & 3.38 & 315 & \textbf{3.49} & 131 \\
\midrule
\textbf{Overall}
            & 2.03 & \textbf{647} & \textbf{2.24} & 280 & 2.15 & 129 \\
\bottomrule

\end{tabular*}
\endgroup
\end{table}

Table~\ref{tab:decomp_per_query} reports compression ratio and decompression throughput per query, aggregated over the columns each query references. Throughout this section, every measurement averages $10$ timed runs after one discarded warmup.

The two schemes trade ratio for speed. RFOR compresses best on six of eight query families and overall ($2.24\times$ vs. $2.03\times$ for FLBP and $2.15\times$ for nvCOMP Cascaded), as RLE captures long runs, notably in \texttt{lo\_orderdate} and  \texttt{lo\_custkey} for SSB Q1.1/Q3.1. FLBP cannot exploit these runs, but its warp-level, synchronization-free decoder is fastest on all
eight families ($559$--$729$\,GB/s). 
  
\subsection{Decompression strategies}
\label{sec:exp-in-gpu-strategies}

\begin{figure}
	\includegraphics[width=\columnwidth]{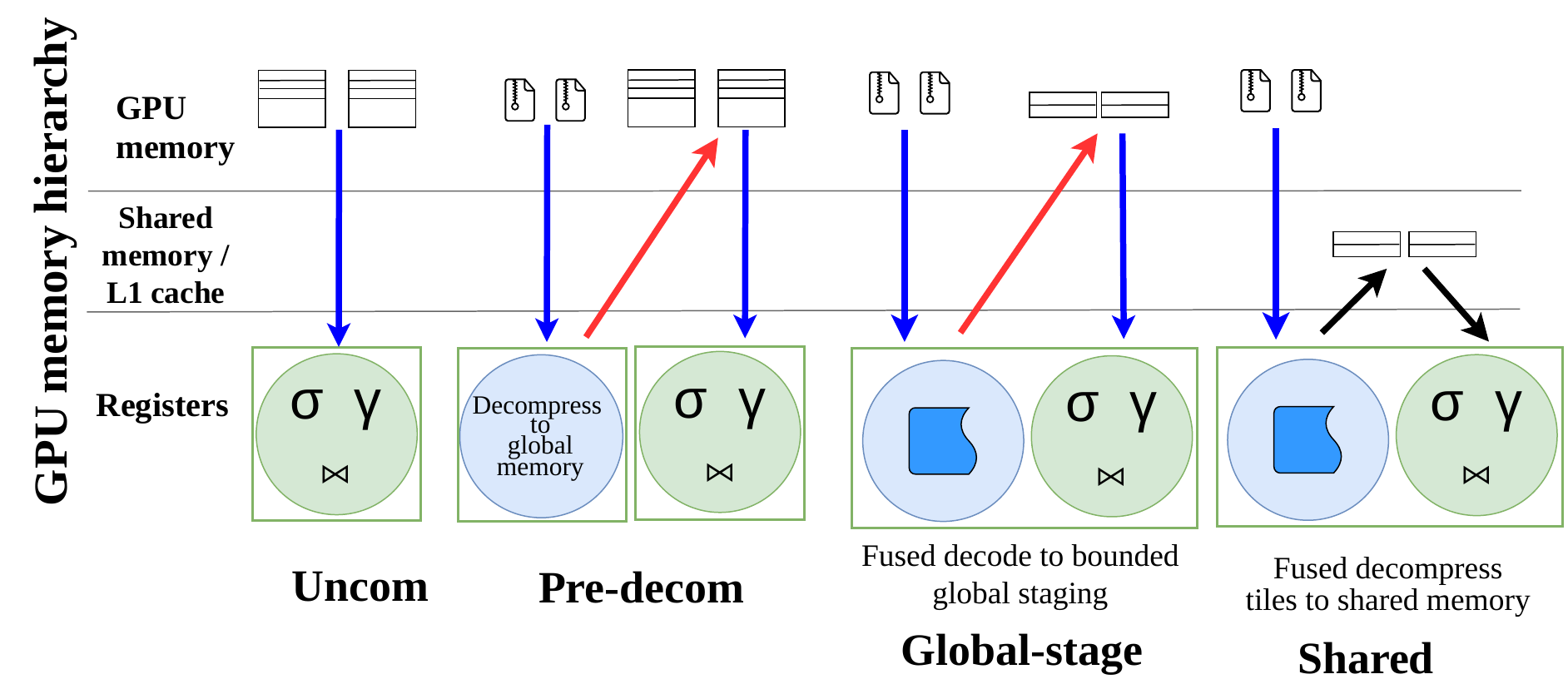}
	\caption{Decompression placement strategies for GPU query execution. 
    Arrow colors correspond to the segments in \Cref{fig:bw}: blue = HBM read, red = HBM write.
	}\label{fig_operator}
\end{figure}

We evaluate the placements in \Cref{fig_operator}:

\noindent{ \bf (\none) Uncompressed baseline.}
In the baseline, the data is stored uncompressed in GPU memory. The \BtTAcro operator has each thread load the tuples it needs and feed the rest of the pipeline.

\noindent{ \bf (\predecom) Decompress to global memory.}
Compressed blocks are decompressed into a preallocated global buffer, after which the unchanged baseline pipeline reads tuples from that buffer. The implementation launches a kernel that performs a cooperative decode using dynamic shared memory
; each block writes its decoded tile directly into a preallocated global memory output buffer. Downstream code is unchanged; each decoded value is written once by the kernel and later read once by the consumer. Thus, it breaks the pipeline: decoded data are first materialized in memory, and only then consumed by a separately launched relational kernel.

 \noindent{\bf (\globalstage) Tile decompress to bounded global memory.}
\ourOpAcro decodes and consumes tiles in the same kernel but stages decoded values in a bounded, reusable global-memory pool. A grid of $W$ worker thread blocks grid-strides over the input; each worker owns one slot containing $T$ four-byte
  values for each of the $K$ projected columns. The pool therefore occupies $4WKT$ bytes, independent of dataset cardinality. Each thread block decodes the columns using one reusable shared-memory tile, copies them into its slot, executes the
  tuple loop, and then reuses the slot. Unlike \predecom, \globalstage neither materializes the complete decoded dataset nor launches a separate consumer kernel.
  
\noindent{\bf (\shared) Tile decompress to shared memory.}
\ourOpAcro cooperatively decodes one tile per projected column into the SM scratchpad and feeds the inner tuple loop directly from there (\Cref{ssec:tile-size-knob}). After a block-wide barrier, the tuple loop reads the decoded values from shared memory. Unlike \globalstage,
\shared neither copies these values to global memory nor reloads them before invoking downstream operators. No decoded HBM buffer is ever materialized. The cost is on-chip capacity: shared-memory usage grows with the projected column count, limiting the strategy's reach.

\begin{figure*}[th]
\centering
    	\includegraphics[width=\textwidth]{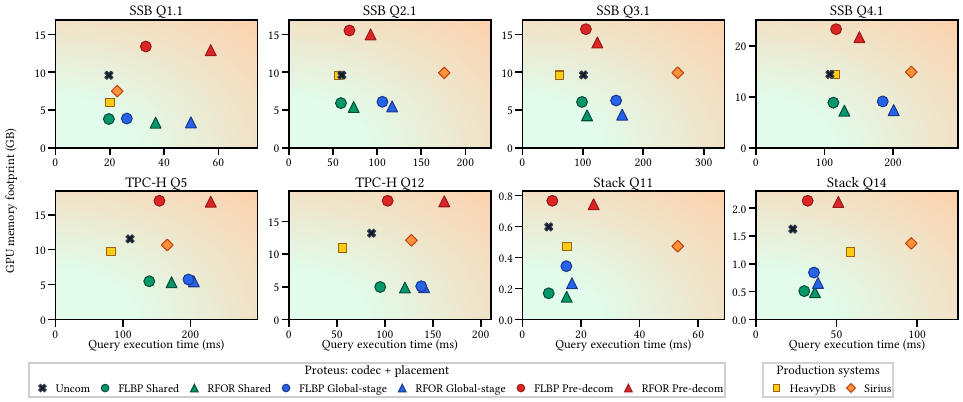}
	\caption{
        GPU memory footprint vs.\ query execution time for SSB Q1.1--Q4.1, TPC-H Q5 and Q12, and Stack Q11 and Q14.}\label{fig:footprint_vs_time}
\end{figure*}

\noindent{\bf (\register) Decompress into registers.}
This variant decodes one column at a time into a reusable shared-memory tile and then moves each thread's values into registers, reducing the shared-memory footprint from $K$ tiles to one at the cost of additional register pressure. FLBP supports direct per-position decoding, while RFOR additionally requires cumulative run boundaries and a per-thread search. Register placement benefits simple queries on the A40, but not on the A100, where its register pressure offsets the shared-memory savings, confirming the findings of \cite{Afroozeh:2024}. We therefore omit it from \Cref{fig_operator,fig:footprint_vs_time}, but further analyze it in \Cref{sec:exp-ssb-e2e}.

\subsection{GPU DRAM traffic decomposition}\label{sec:exp:bw}

\begin{figure}
	\includegraphics[width=\columnwidth]{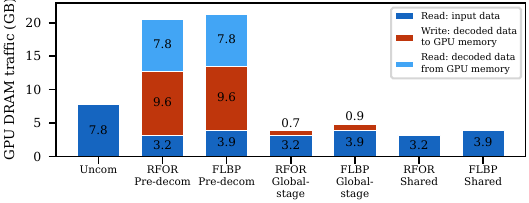}
	\caption{
    GPU DRAM traffic breakdown for Q1.1 decomposed by phase. Segment colors match the arrows in \Cref{fig_operator} (blue = HBM read, red = HBM write). 
	}\label{fig:bw}
\end{figure}

We profile GPU DRAM traffic with NVIDIA Nsight Compute~\cite{nvidia_nsight_compute_2025} (NCU), collecting 
DRAM reads and writes across all kernels in the query. We restrict the analysis to Q1.1, which is scan-dominated and join-free: join queries exchange intermediate results through the GPU cache, and NCU's per-kernel serialization evicts this shared state between replays, generating re-fetches that do not occur during normal execution and would render bandwidth attribution meaningless.

Each bar in \Cref{fig:bw} separates input reads, decoded-value writes, and decoded-value reads. The RFOR and FLBP input segments use their resident compressed sizes, $3.2$ and $3.9$~GB, respectively; the remaining values represent physical HBM traffic measured by NCU. Uncompressed execution reads $7.76$~GB from HBM.

\predecom writes the full $9.60$~GB decompressed dataset into GPU memory buffers and re-reads it from the query kernel. Including the compressed-input read, this produces $20.60$~GB of physical HBM traffic for RFOR and $21.13$~GB for FLBP, respectively $2.65\times$ and $2.72\times$ the uncompressed baseline.

For \globalstage, we size the worker pool to expose all available thread-block parallelism while keeping the staging buffer bounded; for Q1.1, it occupies only $63.7$~MB. \globalstage stores logically and reloads the complete $9.60$~GB decoded stream. Surprisingly, the workers repeatedly access the same portion of the $63.7$~MB array using the GPU cache. NCU reports L2 hit rates of $79.1\%$ for RFOR and $74.2\%$ for FLBP, so only $0.70$ and $0.90$~GB of the logical writes reach HBM. Total physical HBM traffic is $3.98$~GB for RFOR and $4.78$~GB for FLBP.

\shared removes the global-memory handoff: the tuple loop consumes each decoded tile directly from shared memory before the tile is overwritten. Its bars therefore contain only the compressed input: $3.2$~GB for RFOR and $3.9$~GB for FLBP.

The main result is that avoiding whole-dataset materialization and fusing decompression and query execution removes most HBM traffic due to decompression. Relative to \predecom, \globalstage reduces physical HBM traffic by $5.2\times$ for RFOR and $4.4\times$ for FLBP, while \shared reduces it by $6.3\times$ and  $5.6\times$. 

\subsection{Comparing encoders and decompression placements against uncompressed Proteus}\label{sec:exp-ssb-e2e}

\Cref{fig:footprint_vs_time} reports end-to-end query execution time and GPU memory footprint, including the referenced base columns and any placement-specific HBM staging or materialization. 
For Proteus with \ourOpAcro we show the fastest configuration $\tpb \times \ipt$ among: (128$\times$8), (256$\times$4), and (512$\times$2) for FLBP, and (64$\times$8), (128$\times$4) and (256$\times$2) for RFOR.

\noindent{\bf SSB.}
Across Q1.1--Q4.1, \shared provides the best compressed runtime in all eight query--codec cases. On scan-intensive Q1.1, FLBP-\shared matches \none at $19.7$\,ms while reducing footprint from $9.6$ to $3.8$\,GB; RFOR-\shared reaches $3.3$\,GB but takes $36.9$\,ms because its
multi-layer decoder is more expensive. As joins and aggregation add work, decompression is increasingly amortized: on Q2.1--Q4.1, FLBP-\shared ranges from $2\%$ faster to $5\%$ slower than \none, while RFOR's overhead falls from $87\%$ on Q1.1 to $6$--$22\%$. Wider pipelines nevertheless increase resource pressure: for the selected FLBP-\shared kernels, spill stores/loads rise from $44/56$ bytes on Q2.1 to $64/80$ on Q3.1 and $112/136$ on six-column Q4.1. \globalstage retains a footprint close to \shared but is $33$--$80\%$ slower. It publishes and reloads every decoded value through the reusable slot in global address space.

\noindent{\bf TPC-H.}
  Q5 exercises a six-relation pipeline centered on the large SF100 \texttt{lineitem}--\texttt{orders} join, carrying fact tuples through five hash probes before aggregation. FLBP-\shared reduces footprint from $11.53$ to $5.46$\,GB and runs in $139$\,ms versus $110.5$\,ms for \none. RFOR provides similar compression but takes $172$\,ms, consistent with its lower decoder throughput ($284$ versus $653$\,GB/s). Q12 exer\-cises the same large join in a shallower, filter-heavy pipeline. FLBP-\shared obtains a $2.66\times$ footprint reduction at $95.2$\,ms, only $10\%$ above \none; RFOR reaches $2.70\times$ at $121$\,ms. \globalstage takes $197/205$\,ms on Q5 and $138/141$\,ms on Q12 under FLBP/RFOR. Even for these larger and complex join pipelines, eliminating per-column shared staging with \globalstage does not improve runtime.

  \noindent{\bf Stack.}
  Stack's real-world distributions produce the highest compression ratios in the evaluation, ranging from $3.2\times$ to $4.1\times$. On Q11,
  FLBP-\shared reduces footprint from $0.597$ to $0.168$\,GB while matching \none ($8.97$ versus $8.99$\,ms). Wider Q14 scans 21 columns across
  seven relations; repeated decode before its dictionary-ID and range predicates increases FLBP-\shared to $30.0$\,ms and RFOR-\shared to $36.6$
  \,ms, versus $22.9$\,ms for \none, while retaining $3.19\times$ and $3.33\times$ compression. The bounded \globalstage allocation is proportionally visible at this scale: it adds $88$--$175$\,MB on Q11 and $167$--$334$\,MB on Q14.

We profile each configuration with 
NCU, and also report register spill store/load bytes (from \texttt{ptxas} at compile time) in \Cref{tab:rfor-decom-op-q11-q41} as diagnostic evidence.

\noindent{\bf Shared vs Registers}. \shared consistently outperforms \register, which matched it on only a few queries and was never faster. GPU A100 profiles show why: occupancy is constrained by register pressure rather than shared-memory capacity, so moving decoded values into registers does not free additional occupancy on this hardware. \register\ therefore incurs its extra spilling cost with no compensating occupancy gain. Even \globalstage, which reduces the shared-memory footprint by $41$--$83\%$, leaves occupancy unchanged. We therefore use \shared as the default placement.

\noindent{\bf Launch Configuration}. \Cref{tab:rfor-decom-op-q11-q41} sweeps the \tpb/\ipt split of the launch configuration - thread-level parallelism versus work per thread~\cite{volkov2010better} - at fixed tile sizes (512 for RFOR and 1024 for FLBP), on the two SSB endpoints: scan-dominated Q1.1 and join-heavy Q4.1. For RFOR, \ipt$=4$ is a single, query-independent optimum: it wins at \emph{every} query in our suite (SSB, TPC-H, Stack). RFOR's decode is register-hungry, and spilling grows monotonically with \ipt: \ipt$=8$ spills heavily; \ipt$=4$ relieves this pressure by lowering work per thread and raising threads per block, which increases occupancy; \ipt$=2$ gains no further occupancy yet leaves too little work per thread to amortize synchronization. For FLBP, the optimum \ipt\ instead varies between $8$ and $4$ across the same suite. Register pressure explains the SSB queries: Q1.1 spills nothing at any geometry, so more work per thread is effectively free and \ipt$=8$ wins, whereas Q4.1, which spills at every geometry, prefers \ipt$=4$. Unlike RFOR, however, this rule does not generalize beyond SSB: a few queries, e.g., TPC-H Q5 (not shown), spill a high number of registers yet prefer the higher \ipt$=8$ and get slower at the lower-pressure \ipt$=4$, so no single fixed geometry can be recommended the way it can for RFOR. In both codecs, moving past the sweet spot in either direction is never beneficial, consistent with the broader observation that maximizing thread count is not the same as maximizing throughput~\cite{volkov2010better,cuda-best-practices}.

\begin{table}[t]
\centering
\caption{\small\tpb$\times$\ipt sweep for \shared: RFOR (512-value tiles) and FLBP (1024) on SSB Q1.1/Q4.1. Spill S/L: register-spill store/load bytes; bold: fastest per query.}
\label{tab:rfor-decom-op-q11-q41}
\resizebox{\columnwidth}{!}{%
\begin{tabular}{llrrrrrr}
\toprule
& & \multicolumn{3}{c}{Q1.1} & \multicolumn{3}{c}{Q4.1} \\
\cmidrule(lr){3-5}
\cmidrule(lr){6-8}
Codec & Metric & \ipt$=8$ & \ipt$=4$ & \ipt$=2$ & \ipt$=8$ & \ipt$=4$ & \ipt$=2$ \\
\midrule
\multirow{4}{*}{RFOR} & \tpb               & 64 & 128 & 256 & 64 & 128 & 256 \\
                      & Spill S/L (B)      & 136/172 & 44/44 & 8/8 & 152/176 & 112/136 & 100/124 \\
                      & Exec.\ time (ms)   & 49.6    & \bf{36.9}  & 59.6 & 275.0   & \bf{129.0} & 153.0 \\
\cmidrule(lr){1-8}
\multirow{4}{*}{FLBP} & \tpb               & 128 & 256 & 512 & 128 & 256 & 512 \\
                      & Spill S/L (B)      & 0/0 & 0/0 & 0/0 & 156/180 & 112/136 & 100/124 \\
                      & Exec.\ time (ms)   & \bf{19.7} & 24.2 & 34.6 & 150.0 & \bf{119.0} & 151 \\
\bottomrule
\end{tabular}%
}
\end{table}

\noindent{\bf Summary.} Across the eight queries, \shared is the fastest compressed placement in all 16 query--codec cases. Averaging per-query percentages, FLBP-\shared reduces footprint by $54\%$ with only $8\%$ runtime overhead, while RFOR-\shared reduces it by $60\%$ at a $45\%$ overhead. \globalstage retains $46\%$/$56\%$ footprint reductions for FLBP/RFOR but incurs $62\%$/$88\%$ runtime overhead: L2 absorbs most staged traffic, but has higher latency than SM-local shared memory/L1 and global instructions remain. \shared avoids both this handoff and \predecom's full decoded materialization. Our guidance is therefore to decode into shared memory by default. The remaining decision, launch configuration, does not generalize to a single rule: for RFOR, we recommend \tpb$=128$, \ipt$=4$; for FLBP, the optimum varies between $128 \times 8$ and $256 \times 4$; we therefore recommend restricting tuning to these two configurations.
  
\looseness=-1
\subsection{Production GPU database baseline}\label{ssec:competitors}

In \Cref{fig:footprint_vs_time} we also compare Proteus against two external GPU database baselines (described in \Cref{ssec:gpu_execution_models}): Sirius (commit \texttt{8d8353a}, DuckDB~\texttt{v1.5.2}) and HeavyDB (commit \texttt{da23f70}). The two systems bracket the GPU-engine design space and let us anchor Proteus's runtimes against architectural extremes.
For Sirius, we use \texttt{gpu\_processing}, the in-memory path, which matches our GPU-resident setting. We exclude \texttt{gpu\_execution}/Super Sirius because it introduces a different Parquet/task/tiered-memory regime. Both paths use a vectorized, libcudf-based execution model: operators materialize columnar outputs in device memory, whereas compiled engines fuse pipelines so intermediate values flow through registers or shared memory. This trades higher inter-operator memory traffic for avoiding per-query compilation overhead~\cite{Cao:2023,10.14778/3275366.3284966}.
  
The reported intervals for all three systems exclude query compilation and SQL planning; they cover GPU pipeline execution through result materialization, making the comparison independent of each system's SQL front-end overhead.

HeavyDB and Sirius are configured to use the narrowest integer types available in their current implementations. This gives the production baselines their strongest coarse-grained footprint reduction. In \Cref{fig:footprint_vs_time}, for example, in SSB Q1.1 \texttt{lo\_quantity} ranges over $[1,50]$ and \texttt{lo\_discount} over $[0,10]$. HeavyDB stores them as \texttt{INT8}, reducing footprint from $9.6$ to $6.0$\,GB, while Sirius lacks \texttt{INT8} and uses \texttt{INT16}, yielding $7.2$\,GB. Column-level narrowing helps but remains bounded by one physical type per column. Proteus \ourOpAcro instead compresses at tile granularity, exploiting local distributions within each column.

Across the eight queries in \Cref{fig:footprint_vs_time}, fixed-width narrowing reduces HeavyDB's and Sirius's base-column footprints by $15\%$ and $8\%$ on average relative to Proteus-\none. Nevertheless, FLBP-\shared uses $46\%$ less memory than HeavyDB and $50\%$ less than Sirius, while RFOR-\shared uses $53\%$ and $57\%$ less, respectively. HeavyDB is faster than Proteus-\none on four queries and is only $3\%$ slower in geometric mean, whereas Sirius is $2.37\times$ slower. HeavyDB's advantage on TPC-H Q5/Q12 reflects specialized GPU plans that use direct-index perfect-hash joins and keyless aggregations; Proteus uses general build/probe pipelines followed by partial and final aggregation. Sirius's operator-at-a-time execution instead materializes intermediate results between operators, which is particularly costly for the join-heavy queries. Thus, these results compare equivalent queries under each production system's selected physical plan rather than isolating identical join implementations.

\noindent{\bf Summary.}
Across all eight queries, HeavyDB remains runtime-competitive through specialized physical plans, while Sirius is slower than Proteus-\none on every query; neither matches CB2T's footprint. Averaging per-query percentages, FLBP-\shared incurs only $8\%$ runtime overhead over Proteus-\none while using $46\%$ less GPU memory than HeavyDB and $50\%$ less than Sirius. Thus, CB2T combines fused compiled execution with a resident footprint unattainable through fixed-width type narrowing alone.

\looseness=-1
\subsection{Specialized compressed-execution baselines}
  \label{ssec:compressed-execution-baselines}

To isolate CB2T from the general benefit of compressed execution, we compare Proteus with the open-source, tile-based Crystal implementations of GPU-RFOR~\cite{Shanbhag:2022} and FastLanes-GPU~\cite{Afroozeh:2024}. Over SSB (the only available workload for Crystal) the specialized compressed kernels are $2.7$-$4.7\times$ faster than their Proteus counterparts. Most of this gap exists independently of compression: uncompressed Crystal is already $2.0$--$4.1\times$ faster because its queries use fixed, hand-written CUDA execution structures and query-specific data structures. Normalizing each compressed configuration to its own uncompressed baseline therefore better isolates decompression cost. RFOR changes Crystal's runtime by $+31$--$94\%$, compared with $+6$--$87\%$ in Proteus. FLBP changes Crystal's runtime by $-47\%$ to $+66\%$, whereas Proteus remains between $-2\%$ and $+5\%$. Consequently, \ourOpAcro’s advantage is not lower absolute kernel time, but achieving the same footprint reduction with lower RFOR overhead and substantially more stable FLBP performance across query shapes. It thus brings the space benefits of specialized compressed execution into a general compiled engine while keeping composability.

\looseness=-1
\section{Future Work}

This work opens three promising directions. First, compiled GPU engines can make decompression placement hardware- and workload-aware: the compiler already observes the projected columns, operator shape, backend, and target SM resources, but selecting just-in-time the best \texttt{threads\_per\_block}, \texttt{items\_per\_thread}, and decode location requires models that capture the interaction between decompression and the fused query pipeline. 
Second, compression semantics can become compiler-visible. Encoding metadata such as tile minimum value, bit widths, and run boundaries could be represented in the generated LLVM IR and consumed by optimization passes before NVPTX lowering, enabling predicates to skip tiles or avoid decoding values that cannot contribute to the result~\cite{11271872}. Third, as query pipelines increasingly span heterogeneous hardware, the abstraction introduced in this paper can be extended from compressed scans to compressed data exchange. A counterpart operator, TupleToCompressedBlocks (\TtCB), could consume the standard tuple interface produced by a pipeline on one device and emit compressed blocks, making compression a compiler-inserted boundary around pipeline breakers or device crossings~\cite{Chrysogelos:2019}. This would allow intermediate results to be compressed before materialization or transfer to another compute unit, reducing HBM footprint and PCIe/NVLink traffic without manually designing hybrid CPU--GPU data-flow strategies~\cite{10.14778/3749646.3749710}. In such distributed pipelines, the compiler could choose not only where operators run, but also which representation crosses each boundary.

\looseness=-1
\section{Conclusion}

This paper fills the compiled, tuple-at-a-time-over-compressed design space by making decompression a first-class physical operator in a JIT-compiled GPU engine. \ourOpAcro confines encoding-specific logic behind reusable backend hooks, so a new compression format is a single backend rather than a pipeline rewrite, and brings decompression within the compiler's reach, opening automatic optimizations that hand-written kernels cannot express. Executing directly over compressed data reduces both memory traffic and footprint, letting queries fit within smaller GPU-memory budgets and more workloads share the same device.

\begin{acks}
We thank Milos Duric, Charles Benello, and Arturo Paras for their help with preliminary experiments.
This work was supported by the Swiss State Secretariat for Education, Research and Innovation (SERI) for the Prodasys project, a European Research Council (ERC) Advanced Grant.
\end{acks}

\bibliographystyle{ACM-Reference-Format}
\bibliography{references}

\end{document}